\documentclass[format=acmsmall, review=false, screen=true]{acmart}
\usepackage{booktabs}
\usepackage{amsmath,amssymb,amsfonts}
\usepackage{algorithmic}
\usepackage{array}
\usepackage{adjustbox}
\usepackage{color,soul}
\usepackage{glossaries-extra}
\usepackage{rotating,tabularx,siunitx}
\definecolor{Blue}{rgb}{0,0,1}
\usepackage{subfig}

\makenoidxglossaries
\newglossaryentry{iot}{name=IoT,description={Internet of Things}}
\newglossaryentry{f2c}{name=F2C,description={Fog to Cloud}}
\newglossaryentry{http}{name={HTTP},description={Hyper Text Transport Protocol}}
\newglossaryentry{amqp}{name={AMQP},description={Advanced Message Queuing Protocol}}
\newglossaryentry{gds}{name={GDS},description={Global Data Space}}
\newglossaryentry{dds}{name={DDS},description={Data Distribution Service}}
\newglossaryentry{xmpp}{name={XMPP},description={Extensible Messaging and Presence Protocol}}
\newglossaryentry{coap}{name={CoAP},description={Constrained Application Protocol}}
\newglossaryentry{mqtt}{name={MQTT},description={Message Queue Telemetry Transport Protocol}}
\newglossaryentry{rest}{name={REST},description={Representational State Transfer}}
\newglossaryentry{json}{name={JSON},description={JavaScript Object Notation}}
\newglossaryentry{uri}{name={URI},description={Uniform Resource Identifier}}
\newglossaryentry{tls}{name={TLS},description={Transport Layer Security}}
\newglossaryentry{dtls}{name={DTLS},description={Datagram Transport Layer Security}}
\newglossaryentry{ietf}{name={IETF},description={Internet Engineering Task Force}}
\newglossaryentry{quic}{name={QUIC},description={Quick UDP Internet Connections}}
\newglossaryentry{spdy}{name={SPDY},description={Speedy}}
\newglossaryentry{sasl}{name={SASL},description={Simple Authentication and Security Layer}}

\usepackage[ruled]{algorithm2e} 

\SetAlFnt{\small}
\SetAlCapFnt{\small}
\SetAlCapNameFnt{\small}
\SetAlCapHSkip{0pt}
\IncMargin{-\parindent}

\acmJournal{CSUR}
\setcopyright{acmlicensed}

\acmDOI{0000001.0000001}

\begin{document}

\title[Internet of Things Communication Protocols and Fog to Cloud computing]{A Survey of Communication Protocols for Internet of Things and Related Challenges of Fog and Cloud Computing Integration}

\author{Jasenka Dizdarevi{\'{c}}}
\affiliation{%
  \institution{Technische Universit{\"a}t Braunschweig}
  \city{Braunschweig}
  \country{Germany}}
\email{j.dizdarevic@tu-bs.de}

\author{Francisco Carpio}
\affiliation{%
  \institution{Technische Universit{\"a}t Braunschweig}
  \city{Braunschweig}
  \country{Germany}
}
\email{f.carpio@tu-bs.de}

\author{Admela Jukan}
\affiliation{%
 \institution{Technische Universit{\"a}t Braunschweig}
 \city{Braunschweig}
 \country{Germany}}
\email{a.jukan@tu-bs.de}

\author{Xavi Masip-Bruin}
\affiliation{%
  \institution{Universitat Polit\`ecnica de Catalunya}
  \city{Vilanova}
  \country{Spain}
}
\email{xmasip@ac.upc.edu}

\begin{abstract}
The fast increment in the number of IoT (Internet of Things) devices is accelerating the research on new solutions to make cloud services scalable. In this context, the novel concept of fog computing as well as the combined fog-to-cloud computing paradigm is becoming essential to decentralize the cloud, while bringing the services closer to the end-system. This paper surveys on the application layer communication protocols to fulfill the IoT communication requirements, and their potential for implementation in fog- and cloud-based IoT systems. To this end, the paper first briefly presents potential protocol candidates, including request-reply and publish-subscribe protocols. After that, the paper surveys these protocols based on their main characteristics, as well as the main performance issues, including latency, energy consumption and network throughput. These  findings are thereafter used to place the protocols in each segment of the system (IoT, fog, cloud), and thus opens up the discussion on their choice, interoperability and wider system integration. The survey is expected to be useful to system architects and protocol designers when choosing the communication protocols in an integrated IoT-to-fog-to-cloud system architecture.
\end{abstract}

%
%



\maketitle

\newpage

\section{Introduction}

Continuous innovations in hardware, software and connection solutions in the last decade have lead to the expansion of the Internet of Things (IoT) with the  number of connected devices growing by the day \cite{Hahm2016} \cite{Xu2017}. The huge amount of data generated by these devices require to find a proper system architecture able to both process and store all the data. While cloud-based architectures are being currently used for that purpose, the new fog computing paradigm is envisioned to scale and optimize the \gls{iot} infrastructures \cite{Sethi2017}. Examples of the cloud-based IoT solutions have been proposed in \cite{Botta2014}, \cite{Huo2014}, \cite{Fortino2014} and a detailed analysis of properties for IoT cloud providers has been conducted in \cite{7522237}. These studies have shown that cloud computing has the potential to satisfy many IoT requirements, such as monitoring of services, powerful processing of sensor data streams and visualization tasks. On the other hand, fog-based solutions are suited to address real-time processing, fast data response, and latency issues, thus extending the cloud capabilities closer to the edge of the network \cite{Bonomi2012}. Among many factors that will determine the performance in a combined IoT, fog and cloud computing paradigm, the application layer communication, which in turn depends on the selected communication protocols, is one of the main ones.

Despite the popularity and wide spread usage of HTTP, the currently used protocols in various domains of IoT, fog and cloud domains are de-facto fragmented with many different solutions. This is due to the different requirements and areas that IoT needs to cover, combining the functionalities of sensors, actuators and computing power with security, connectivity and a myriad of other features. As a result, there is no common agreement on the reference architecture or adopted standards of communication protocols. Thus, one of the fundamental challenges for system engineers is to choose the appropriate protocol for their specific IoT system requirements, while levering the advances in fog and cloud computing.  For this challenge to be addressed, some general architecture requirements need to be taken into consideration. These requirements include: devices that can range from resource constrained devices to high performance cloud systems,  data generated to be processed between the cloud and the fog layer, the types of wireless connectivity that can be used, or security and privacy solutions, just to name a few. While there have been several surveys covering different aspects of the IoT architecture \cite{Fortino2014, Marjani2017, Babu2015, Hahm2016, Clarke, Solapure2016, Granjal2015}, the specific issues of communication protocols in the application layer have not been addressed yet.

This paper surveys communication protocols in the application layer (also referred to as messaging protocols and machine-to-machine, depending on the context) in IoT architectures in the context of specific challenges in fog and cloud computing integration, including \gls{mqtt} (Message Queuing Telemetry Transport), \gls{amqp} (Advanced Message Queuing Protocol), \gls{xmpp} (Extensible Messaging and Presence Protocol), \gls{dds} (Data Distribution Service), \gls{http} (Hypertext Transfer Protocol) and \gls{coap} (Constrained Application Protocol). Recognizing the fact that one single messaging protocol will not be enough to cover the entire communication on the combined IoT-F2C architecture built by bringing together IoT, fog and cloud systems, our goal is to unveil open issues and challenges towards the end goal: their seamless interoperability, coordination and integration. To this end, the paper first presents a comparative analysis of the main characteristics of IoT communication protocols, including request-reply and publish-subscribe protocols. After that, the paper surveys each protocol in detail, and discusses their implementation in various segments of the system (IoT, fog, cloud), and thus opens up the discussion on their interoperability and wider system integration. Finally, we also review the main performance issues, including latency, energy consumption and network throughput. Compared to other related surveys, including \cite{Al-Fuqaha2015, Foster2014a, Naik2017, Ramirez2013, Swamy2017, Yassein2016, Nastase2017, Karagiannis2015, Masek}, our focus is on communication protocols in the application layer, with the goal of both exploring their current status, as well as exploring the potential for their integration in the combined IoT, fog and cloud systems.

The rest of this paper is organized as follows. Section 2 presents the background of IoT-F2C architectures and protocols based on publish-subscribe and request-reply interaction model. Section 3 gives a detailed overview of the main features of application layer protocols. Section 4 presents a comparative performance analysis of the protocols surveyed. Section 5 presents possible implementation solutions, such as solutions based on a single communication protocol or a combination of protocols, as well as open issues and challenges. Section 6 concludes the paper.
\section{Background}

This section first provides a background on communication protocols for Internet of Things (IoT) and related scenarios towards fog and cloud computing integration, which motivates the survey. We also provide a brief introduction to communication protocols as basis for more detailed descriptions in the following sections.

\begin{figure}[t]
	\centering
	\includegraphics[width=0.6\columnwidth]{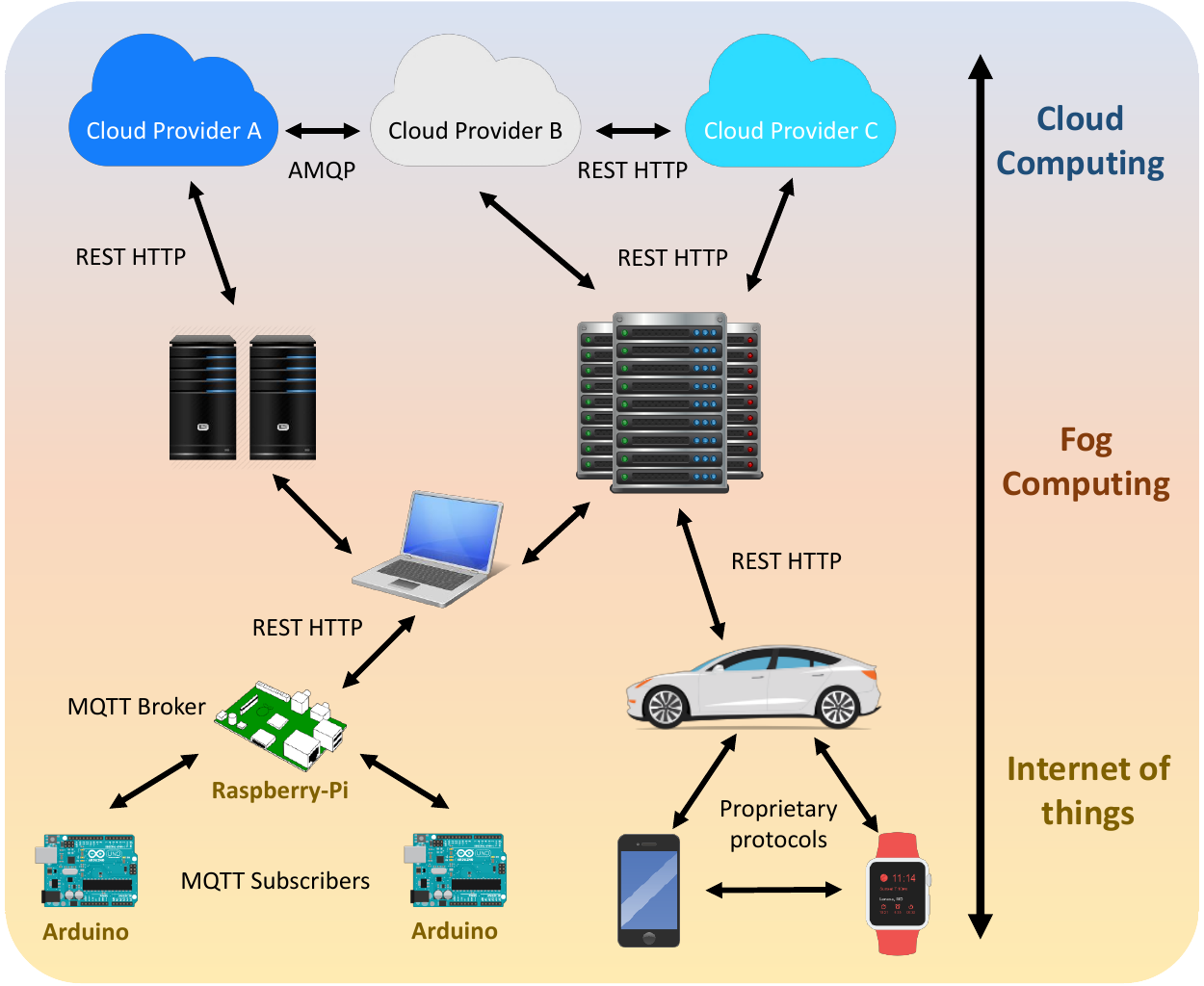}
    \caption{IoT and Fog to Cloud systems}
    \label{arch1}
\end{figure}

\subsection{A Fog-to-Cloud Architecture (F2C) for IoT}

Recently, notable efforts have been devoted to analyze the advantages and benefits brought by an efficient and coordinated management of IoT, cloud and fog. A few standardization initiatives and industrially led researach consortia highlight their importance, such as the OpenFog Consortium  \cite{opencons}, Edge Computing Consortium \cite{ecc} and the mF2C H2020 EU project \cite{mF2C}. While previous cloud-based solutions only consider two layers, the cloud and the IoT end-devices, these recently proposed combined IoT-fog-to-cloud systems introduce new functional abstractions in between. These abstractions can include a single fog computing layer, whereby the fog computing layer itself can be divided into multiple abstraction sub-layers, depending on various factors, such as resource specifications or set of policies defined to accommodate the different devices into layers. We illustrate one such abstraction with Fig.\ref{arch1} along with the candidate communication protocols. In this typical IoT-fog-cloud ecosystem, the \gls{iot} devices are positioned to send data to more capable servers and computing systems in the fog computing layer, such as to perform computing tasks that require low latency. In the same system, cloud computing performs tasks that require larger amounts of computing or storage resources.
As we can see at the bottom of Fig.\ref{arch1}, some IoT devices can be implemented as low-cost processing platforms, such as Arduinos and Raspberry Pis, with MQTT protocol as a communication protocol of choice, as it is optimized to work on constrained devices. It should be noted that devices without computing capabilities are not taken into consideration in regards to communication protocols, since they communicate at the level of hardware that typically does not require interoperability features. Other smart objects, such as smart phones and smart watches, can be considered IoT devices as well in the context of communication protocols. In that case, however, proprietary communication protocols from major vendors are typically implemented. The  IoT data generated is communicated with the fog abstraction layer commonly using the \gls{rest} \gls{http} protocol, which provides flexibility and interoperability for developers to create RESTful (Representational State Transfer) web services. The latter is critical to remaining backwards compatible with the existing computing infrastructure, running on local computers, servers, or cluster of servers. The local resources are commonly referred to as fog nodes \cite{fognode} and are able to filter the received data to be either consumed locally or forwarded to cloud for further computations. While the cloud is usually perceived as a unique entity, in reality, cloud services can use more than one cloud provider to meet requirements for reliability, scalability and economics. For this reason, clouds usually support different communication protocols, with AMQP and REST HTTP being among the most used ones.  Since HTTP is widely accepted and compatible with the current Internet, the natural question would be whether to use HTTP in the IoT and fog layers. However, this protocol, despite its popularity, has been shown to exhibit a lot of performance issues when used in constrained nodes, as we will discuss later in the survey.

\subsection{On Communication Protocols}

\begin{figure}[t]
	\centering
	\includegraphics[width=0.5\columnwidth]{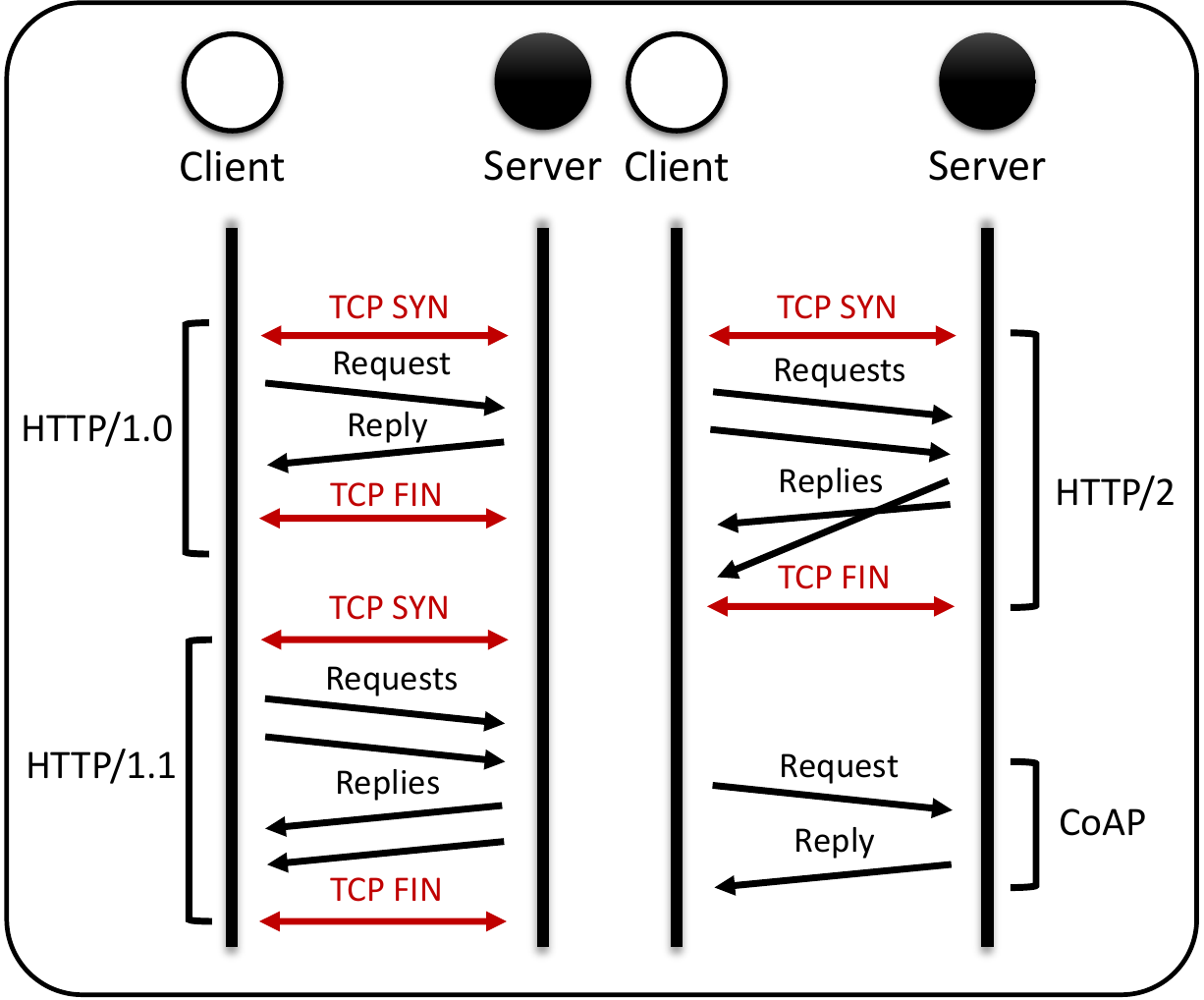}
    \caption{Request-Reply model, for example: COAP and HTTP}
    \label{request-reply}
\end{figure}

In general, the candidate communication protocols differ in their interaction models, i.e., request-reply and publish-subscribe. The request-reply communication model is one of the most basic communication paradigms. It represents a message exchange pattern especially common in client/server architectures. It allows a client to request information from a server that receives the request message, processes it and returns a response message. This kind of information is usually managed and exchanged centrally. The two most known protocols based on the request/reply model are REST HTTP and CoAP.  Fig.\ref{request-reply} shows examples of different client/server interactions, for three HTTP versions (i.e., v.1.0, v.1.1 and v.2.0) as well as for CoAP. In HTTP 1.0, the TCP connection is closed after a single HTTP request/reply pair. In HTTP 1.1, a keep-alive-mechanism was introduced, where a TCP connection could be reused for sending multiple requests to the server without waiting for a response (pipelining). Once the requests are all sent,  the browser starts listening for responses and  HTTP 1.1 specification requires that a server must send its responses to those requests in the same order that the requests were received. The new HTTP 2.0 introduces a multiplexing method by which multiple HTTP requests can be sent and responses can be received asynchronously via a single TCP connection. The fourth  interaction shown is for CoAP, and unlike the others it does not depend on an underlying reliable TCP connection to exchange request/reply messages between the client and the server. The publish-subscribe model, on the other hand, emerged out of the need to provide a distributed, asynchronous, loosely coupled communication between data generators and destinations. The solution appears today in the form of numerous publish-subscribe Message-Oriented Middlewares (MoM) \cite{Jia2014} and recently has been a subject of numerous research efforts \cite{Antonic2015,Chelloug2017,Azzara2013, Veeramanikandan2017}.

\begin{figure}[t]
	\centering
	\includegraphics[width=0.5\columnwidth]{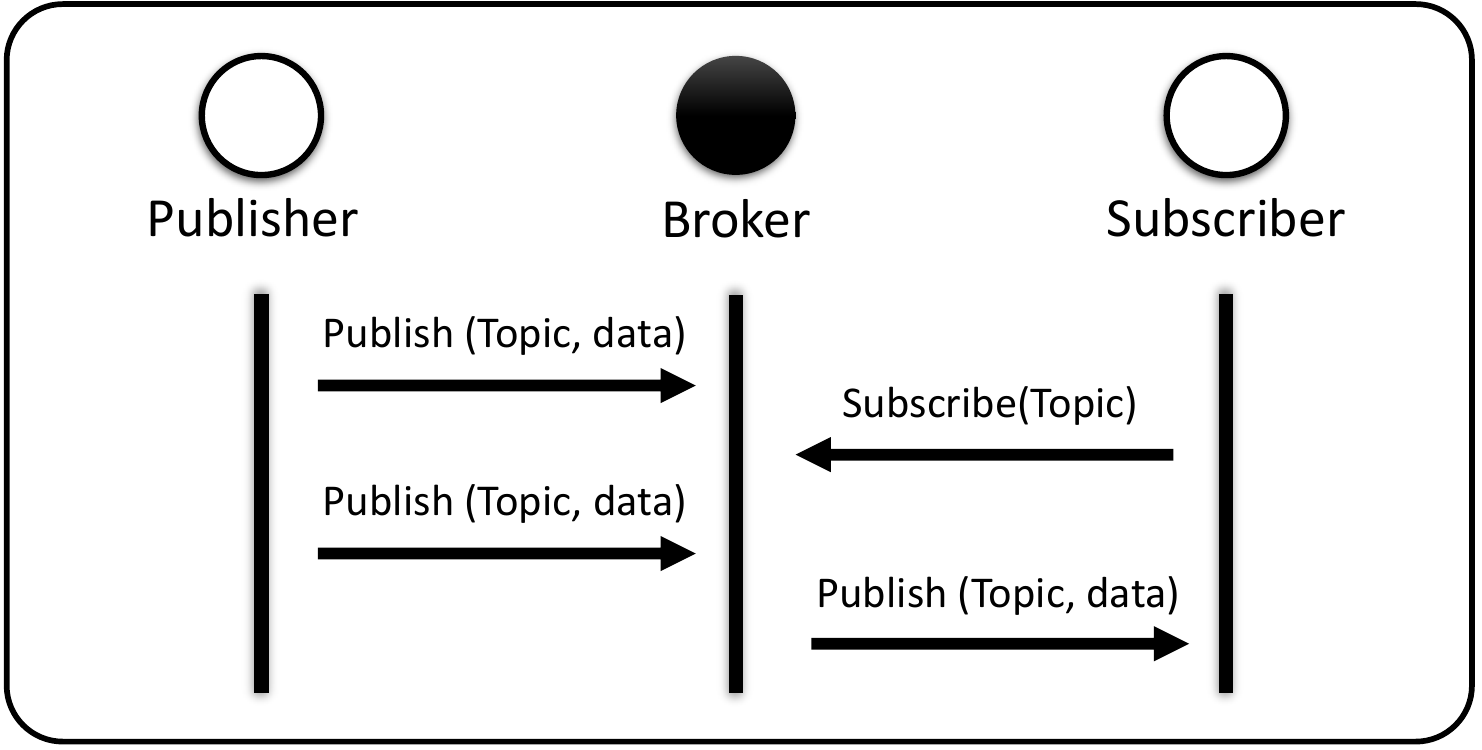}
    \caption{Publish-Subscribe model, for example: MQTT, DDS and AMQP}
    \label{publish-subscribe}
\end{figure}

In this survey, of particular interest are the  protocols based on the publish-subscribe interaction model as an alternative to the traditional request-reply (client-server) model. We can see an example of this interaction model, that consists of three parties, publisher, subscriber and a broker presented in Fig. \ref{publish-subscribe}. Here, the client with a role of a subscriber does not have to request information from the server. Instead of the request, the subscriber interested in receiving messages will subscribe to particular events (topics) within the system. The client subscribes to the broker, the central point in this architecture, responsible for filtering all incoming messages and routing them accordingly between publishers and subscribers \cite{Banavar}. The third party is the publisher that serves as the information provider. When an event about a certain topic occurs, it publishes it to the broker who sends the data on the requested topic to the subscriber. For these reasons, publish-subscribe interaction model can be described as an event-based architecture \cite{Hinze2009}.
This interaction model is interesting for the applications of IoT, fog and cloud computing systems due to its ability to provide scalability and simplify interconnections between different devices, by supporting dynamic, many-to-many and asynchronous communication \cite{10.1007/978-981-10-2750-5_30}.

Comparing the two basic models, i.e., request-reply and publish-subscribe, we can observe that the publish-subscribe model has many benefits: i) publishers and subscribers do not need to know about the existence of each other; ii) one subscriber can receive information from many different publishers and one publisher can send data to many different subscribers (many-to-many communication is supported); iii) publisher and subscriber do not need to be active at the same time to exchange information, because the broker (working as a sort of queuing system) can store messages for clients that are not currently connected \cite{Sachs2010}. There are many standardized messaging protocols currently implementing a publish/subscribe interaction model, most notably MQTT, AMQP and DDS. However, request-reply model also has some advantages. In cases where the capacity of the server side for processing multiple clients requests is not an issue it makes more sense to use already proven and reliable request-reply interactions. So, the choice of the model depends on the application scenario for which it will be used. Finally, some protocols support both request-reply and publish-subscribe interaction models. This includes XMPP protocol, and the new version of the HTTP - HTTP2.0, which supports the server push option, as discussed in Section \ref{http_subs}. \gls{ietf} has also released a draft describing a Publish-Subscribe Broker for other protocols of interest, such as CoAP \cite{pubsubCoap}. In an attempt to solve the message exchange, over the time few other solutions emerged, such as the WebSockets protocol  \cite{websocket} or using HTTP over \gls{quic} (Quick UDP Internet Connections) protocol. In case of WebSocket, although it is used for real-time pushing of data from a server to a web client and enable persistent connections with simultaneous bidirectional communication, it is not designed for resource constrained devices \cite{Karagiannis2015}. QUIC is also rather noteworthy, as a novel transport protocol creating a wave of the new research efforts \cite{quic1, quic2, quic3}. Since QUIC has not been standardized yet, it maybe too early to predict its possible application and impact in IoT based solutions. For these reasons, and despite their novelty, WebSockets and QUIC are out of the scope in this survey.

\begin{table*}[]%
\caption{Application layer protocols main features comparison}
\label{comparison1}
\begin{center}
\setlength\tabcolsep{3pt}
\begin{tabular}{|c|c|c|c|c|c|c|}
  \hline
\textbf{Protocol} & \textbf{Req.-Rep.} & \textbf{Pub.-Sub.} & \textbf{Standard} & \textbf{Transport} & \textbf{QoS}                & \textbf{Security}     \\ \hline
REST HTTP         &  \checkmark            &                            & IETF \cite{Fielding1999}& TCP                         & -                           & TLS/SSL               \\ \hline
MQTT              &                        &   \checkmark               & OASIS \cite{OASIS2014}  & TCP                         & 3 levels                    & TLS/SSL               \\ \hline
CoAP              &   \checkmark           &   \checkmark               & IETF \cite{Shelby2014} & UDP                         & Limited                     & DTLS                  \\ \hline
AMQP              &   \checkmark           &   \checkmark               & OASIS \cite{OASIS2012}  & TCP                         & 3 levels                    & TLS/SSL               \\ \hline
DDS               &            &       \checkmark                        & OMG \cite{OMG2015}      & TCP/UDP                     & Extensive 					& TLS/DTLS/DDS sec. \\ \hline
XMPP              &     \checkmark         &   \checkmark               & IETF \cite{Saint-andre2017}& TCP                         & -                          & TLS/SSL               \\ \hline
HTTP/2.0          &     \checkmark         &   \checkmark               & IETF \cite{Thomson2015}& TCP                         & -                           & TLS/SSL               \\ \hline
\end{tabular}
\end{center}
\end{table*}

\section{Communication protocols overview}

This section describes the above mentioned protocols based on their main features, as summarized in Table \ref{comparison1}. In a nutshell, Table \ref{comparison1}  summarizes the  standardization status, interaction model, quality of service options, transport protocol and security mechanisms. MQTT, AMQP, XMPP and REST HTTP, are designed to run on networks that use TCP, while CoAP uses UDP as the underlying transport. DDS primarily uses UDP as its underlying transport, but it also supports TCP. As mentioned in the previous section, MQTT, AMQP and DDS implement a publish/subscribe model, while REST HTTP and CoAP implement a request/reply interaction model. MQTT, AMQP and CoAP protocols provide very basic QoS support for delivering messages. MQTT and AMQP implement three different QoS levels, while in CoAP request and reply messages are limited to two. The QoS in REST HTTP and XMPP  is provided by the underlying transport protocols. DDS, on the other hand, provides a rich set of QoS policies with over 20 different QoS options defined by the standard \cite{Foster2014a}. Most of these protocols choose TLS or DTLS protocol as security mechanisms.  Readers interested more in applications of these protocols in various segments (IoT, fog, and cloud) and less so in the protocols design itself, or readers familiar with individual protocols, can skip this section or parts of it, and use the overview in Table \ref{comparison1} to follow up on further discussions in Section \ref{section4}.

\subsection{Hyper Text Transport Protocol (HTTP)} \label{http_subs}
\begin{figure}[t]
	\centering
	\includegraphics[width=1.0\linewidth]{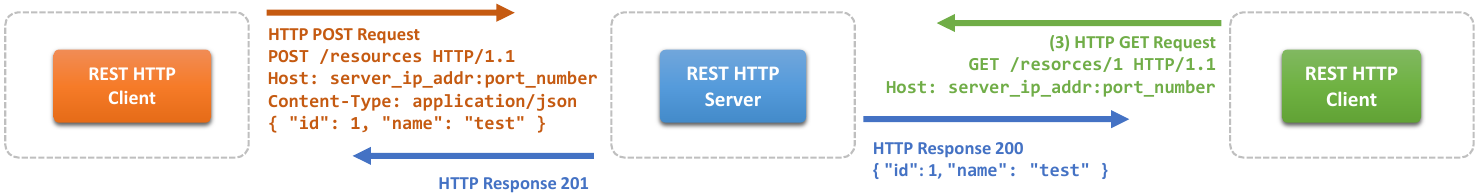}
    \caption{REST HTTP interaction model}
    \label{http_generic}
\end{figure}

This protocol is the fundamental client-server model protocol used for the Web, and the one most compatible with existing network infrastructure, used by the web developers on a daily basis. Currently, the most widely accepted version of this protocol is HTTP/1.1. Communication between a client and a server occurs via a request/response messaging, with client sending an HTTP request message and server then returning a response message, containing the resource that was requested in case the request was accepted. Recently, HTTP has been associated with \gls{rest} \cite{Severance2015}, a guideline for developing web services based on a specific architectural style in order to define the interaction between different components. Because of the success of RESTful Web services, there has been a lot of effort in bringing this architecture into IoT based systems by combining HTTP and REST. The combination of HTTP protocol with REST is commendable, because the devices can make their state information easily available, due to a standardized way to create, read, update, and delete data (the so-called CRUD operations). According to this mapping, the operations for creating, updating, reading and deleting resources correspond to the HTTP POST, GET, PUT and DELETE methods, respectively. For developers, the fact that REST establishes a mapping of these CRUD operations with HTTP methods, means that they can easily build a REST model for different IoT devices \cite{Babovic2016}. The presentation of the data is not pre-defined and as such, the type is arbitrary, with the most common being JSON and XML. In most cases, IoT standardizes around \gls{json} over HTTP. Fig. \ref{http_generic} illustrates an example of a REST HTTP request/reply interaction between two clients and one server . First one of the REST HTTP client wants to add a resource on a server side. For this it is necessary to specify, in the header of the POST method, the root of the resource that will be added \emph{/resources}, the HTTP version, the \emph{Content-type}, which in this case is a JSON file that represents a specific resource, and finally the JSON object itself. The response from the server specifies whether the request was successful, by specifying the HTTP standard status codes (e.g., 201, resource created). For the second client to get this new resource, the GET method has to be specified with the specific URI (e.g. \emph{/resources/1}), which contains the root of the resource and the id of the resource itself. The server will return the JSON object representing the resource. It is worth to mention that beside the simple communication which REST HTTP offers it also has the abundant support and available frameworks making it a default way of web communication, and all servers and client side drivers support it.

Regarding the transport protocol used, HTTP uses TCP. While using TCP provides reliable delivery of large amounts of data which is an advantage in connections that do not have strict latency requirements, it creates challenges in resource constrained environments \cite{Shang2016}. One of the main problems is that the constrained nodes most of the time send small amounts of data sporadically and setting up a TCP connection takes time and produces unnecessary overhead. For QoS, HTTP does not provide additional options, but instead it relies on TCP, which guarantees successful delivery as long as connection is not interrupted.

HTTP as a security mechanism uses the very well-known \gls{tls} \cite{Dierks2008} for enabling secure encrypted communication channel, resulting in a secure version of HTTP, also known as HTTPS.  The first part of securing the client-server data exchange is a TLS handshake, implemented as an exchange of a 'client hello' and a 'server hello' messages where they have to agree upon a cipher suite, which is a combination of algorithms they will use to assure secure settings. After that, the client and server side exchange keys based on the agreed key exchange algorithm. The result is an exchange of messages encrypted with a shared secret key. The data is encrypted to prevent anyone from listening to and understanding the content. In systems that will include resource constrained nodes, current TLS implementation (TLS version 1.2) through its handshake process adds additional traffic with each connection establishment that can deplete the computing capabilities of these devices. Efforts are being made in developing a new TLS version 1.3 that will make TLS handshake faster and lighter, as more convenient for IoT \cite{7958594, 7546519}.

While in general, HTTP presents one the most stable protocol options, there are still a few issues that have lead to the exploration of alternative protocol solutions, due to HTTP complexity, long header fields and high power consumption. Furthermore, HTTP uses the request/reply paradigm, which is not suitable for push notifications, where the server delivers notifications to the client without a client request. Moreover, the TCP protocol overhead maybe too large (three way handshake), especially in case of simple computing nodes in IoT architectures \cite{Savolainen2014}. HTTP does not explicitly define QoS levels and requires additional support for it. This has led to modifications and extension of HTTP, most notably in form of HTTP/2.0 \cite{Thomson2015}, that introduced a number of improvements, some of which are especially relevant in IoT context. HTTP/2.0 enables a more efficient use of network resources and a reduced latency by introducing  compressed headers, using a very efficient and low memory compression format, as well as allowing multiple concurrent exchanges on the same connection \cite{Stenberg}. These features are particularly interesting for the IoT as it means the size of packets is significantly smaller, making it a more adequate option for constrained devices. Additionally, it introduces the so-called server push, which means the server can send content to clients with no need to wait for their requests. The drawbacks of this version of the protocol in IoT based systems are not known yet, as to the best of our knowledge there are no implemented and tested solutions reported in the literature, as of today. It is however likely that one of the drawbacks will be the same as found in HTTP 1.1, i.e., the utilization of TLS protocol as the security mechanism.

\subsection{Constrained Application Protocol (CoAP)}
This protocol was designed by the Constrained RESTful Environments (CoRE) working group of IETF \cite{Shelby2014} for the use in constrained devices with limited processing capabilities. Similar to HTTP, one of its most defining characteristics is its use of tested and well accepted REST architecture. With this feature CoAP supports request/response paradigm just like REST HTTP, and especially so for constrained environments.  CoAP is considered a lightweight protocol, so the headers, methods and status codes are all binary encoded, thus reducing the protocol overhead in comparison with many protocols. It also runs over less complex UDP transport protocol instead of TCP, further reducing the overhead. When a CoAP client sends one or multiple CoAP requests to the server and gets the response, this response is not sent over a previously established connection, but exchanged asynchronously over CoAP messages. The price paid for this reduction is reliability. It should be noted that because of the reduced reliability features, which is known when using UDP,  IETF has created an additional standard document, opening up the possibility of CoAP running over TCP \cite{Lemay2018}. However, at this moment this feature is still in its early stages.

CoAP relies on a structure that is divided into two logically different layers. One of the layers, the so-called request/response layer, implements RESTful paradigm and allows for CoAP clients to use the HTTP-like methods when sending requests. In other words, clients can use GET, PUT, POST or DELETE methods to manage the URI identified resources in the network \cite{Nguyen2016}. Just like in HTTP, for its requests for obtaining data from the server, for instance when obtaining the sensor values, client will use method GET with a server URL, and as a reply will receive a packet with that data.
The request and responses are matched through a token; a token in the response has to be the same as the one defined in the request. It is also possible for a client to push data, for example updated sensor data, to a device by using method POST to its URL. As we can see, in this layer CoAP uses the same methods as REST HTTP. What makes COAP different from HTTP is the second layer. Because UDP does not ensure reliable connections, CoAP relies on its second structural layer for reliability, called the message layer, designed for retransmitting lost packets. This layer defines four types of messages: CON (Confirmable), NON (non-confirmable), ACK (Acknowledgement), and RST (reset). The CON messages are used for ensuring reliable communication, and they demand an acknowledged from the receiver side with an ACK message. Precisely this feature that marks whether the messages need the acknowledgement is what enables QoS differentiation in CoAP, albeit in a limited fashion.

CoAP has an optional feature that can improve the request/response model by allowing clients to continue receiving changes on a requested resource from the server \cite{Correia2016} by adding an {\em observe} option to a GET request. With this option, the server adds the client to the list of observers for the specific resource, which will allow the client to receive the notifications when resource state changes. Instead of relying on repetitive polling to check for changes in resource state, setting an {\em observe} flag in a CoAP client's GET request, allows an interaction much closer to a publish-subscribe paradigm with a server alerting a client when there are changes. In an attempt to get even closer to publish/subscribe paradigm, IETF has recently released the draft of Publish-Subscribe Broker that extends the capabilities of CoAP for supporting nodes with long interruptions in connectivity and/or up-time \cite{pubsubCoap}, with preliminary performance evaluations showing promising results \cite{Huang}.

As a security mechanism CoAP uses \gls{dtls} \cite{Rescorla2012} on top of its UDP transport protocol. It is based on TLS protocol with necessary changes to run over an unreliable connection. The result is a secure CoAPS protocol version. Most of the modifications in comparison to TLS include features that stop connection termination in case of lost or out of order packets. As an example, there is a possibility to retransmit handshake messages. Handshaking process is very similar to the one in TLS, with the exchange of client and server 'hello' messages, but with the additional possibility for a server to send a verification query to making sure that the client was sending its 'hello' message from the authentic source address. This mechanism helps prevent Denial-of-Service attacks. Through these messages, client and server also exchange supported cipher suits and keys, and agree on the ones both sides support, which will further be used for data exchange protection during the communication.

Since DTLS was not originally designed for IoT and constrained devices, new versions optimized for the lightweight devices have emerged recently  \cite{Kumar2015, Raza2013}. Some of the DTLS optimization mechanisms with a goal of making it more lightweight include IPv6 over Low-power Wireless Personal Area Network (6LoWPAN) header compression mechanisms to compress DTLS header \cite{6227754}. Because of its limitations, optimizing DTLS for IoT is still an open issue \cite{Granjal2015,7103240}.

\subsection{Message Queue Telemetry Transport Protocol (MQTT)} \label{mqtt_subs}
MQTT is one of the lightweight messaging protocols that follows the publish-subscribe paradigm, which makes it rather suitable for resource constrained devices and for non-ideal network connectivity conditions, such as with low bandwidth and high latency. MQTT was released by IBM, with its latest version MQTT v3.1 adopted for IoT by the OASIS \cite{OASIS2014}. Because of its simplicity, and a very small message header comparing with other messaging protocols, it is often recommended as the communication solution of choice in IoT. MQTT runs on top of the TCP transport protocol, which ensures its reliability. In comparison with other reliable protocols, such as HTTP, and thanks to its lighter header, MQTT comes with much lower power requirements, making it one of the most prominent protocol solutions in constrained environments.

\begin{figure}[t]
 	\centering
	\includegraphics[width=1.0\linewidth]{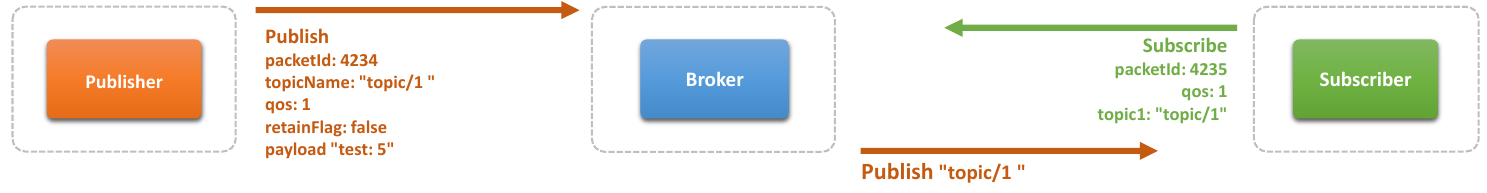}
 \caption{MQTT interaction model}
  \label{mqtt_generic}
 \end{figure}

There are two communication parties in MQTT architecture that usually take the roles of publishers and subscribers, clients and servers/brokers. Clients are the devices that can publish messages, subscribe to receive messages, or both. The client must know about the broker that it connects to, and for its subscriber role it has to know the subject it is subscribing to. A client subscribes to a specific topic, in order to receive corresponding messages. However, other clients can also subscribe to the same topic and get the updates from the broker with the arrival of new messages. Broker serves as a central component that accepts messages published by clients and with the help of the topic and filtering delivers them to the subscribed clients. In MQTT, a publish-subscribe interaction model can be used as illustrated in Fig. \ref{mqtt_generic}. The communication takes place between a broker and two MQTT clients, a publisher and a subscriber. For a device to have a role of the broker, it is necessary to install MQTT broker library, for example Mosquitto broker \cite{mosquitto}, which is one of best known open source MQTT brokers. It should be noted that there are various other MQTT protocol brokers that are open for use, which differ by way of implementation of the MQTT protocol. Some of them are Emqttd \cite{emq2.0}, ActiveMQ \cite{ActiveMQ}, HiveMQ \cite{hivemq}, IBM MessageSight \cite{Maynard2015}, JoramMQ \cite{JoramMQ}, RabbitMQ \cite{rabbitmq}, and VerneMQ \cite{verne}. The clients are realized by installing MQTT client libraries. The publisher creates labeled topics into the Broker, as shown in Fig. \ref{mqtt_generic}. Topics in MQTT are treated as a hierarchy, with strings separated by slashes that indicate the topic level \cite{Tantitharanukul2017}. One MQTT publisher can publish messages to defined set of topics. In this case client will publish the topic: {\em topic/1}. This information will be published to the broker which can temporally store it in a local database. The subscriber interested in this topic sends a subscribe message to a broker, specifying the same topic.

For QoS, MQTT defines three QoS levels, QoS 0, 1, and 2 \cite{OASIS2014,Luzuriaga2015a}. The choice of the level can be defined both in the publish and the subscribe message body. QoS 0 delivers on the best effort basis, without confirmation on message reception. This is a choice in cases where some sensors gather telemetry information over a longer time period, and where the sensors values do not change significantly. It is then acceptable if sometimes the messages are missing, because the general sensor value is still known since most of the message updates have been received. The next level of guarantee is QoS 1, which assures that messages will arrive, so a message confirmation is necessary. This means that receiver must send an acknowledgment, and if it does not arrive in a defined period time, publisher will send a publish message again. The third option, QoS 2, guarantees that the message will be delivered exactly once without duplications. The amount of resources necessary to process MQTT packet increases with the higher chosen QoS level, so it is important to adjust the QoS choice to specific network conditions.

Another important feature MQTT offers is the possibility to store some messages for new subscribers by setting a 'retain' flag in published messages. If there is nobody interested in a topic on which the publisher sends the updates, broker will discard the published messages. But, in some situations, especially when the state of the followed topic does not change often, it is useful to enable for new subscribers to receive the information on that topic.  In this default case new subscribers would have to wait for the state to change in order to receive a message about the topic. By setting a 'retain' flag to value: {\em true}, broker is informed that it should store the published message, so it could be delivered to new subscribers.

MQTT uses TCP which can be critical for constrained devices. To this end, a solution has been proposed as MQTT for Sensor Networks (MQTT-SN) version that uses UDP and supports topic name indexing \cite{Stanford-Clark2013}. This solution does not depend on TCP, but instead uses UDP as faster, simpler, and more efficient transport option over a wireless link \cite{Govindan2015}. The other important improved feature is the reduced size of the payloads. This is done by numbering the data packets with numeric topic \textit{id's} rather than long topic names. The biggest disadvantage is that at the moment MQTT-SN is only supported by a few platforms, and there is only one free broker implementation known, called Really Small Message Broker \cite{Xu2017}.

Since it was designed to be as lightweight, MQTT does not provide encryption, and instead, data is exchanged as plain-text, which is clearly an issue from the security standpoint. Therefore, encryption needs to be implemented as a separate feature, for instance via TLS, which on the other hand increases overhead. Authentication is implemented by many MQTT brokers, through one of the MQTTs control type message packets, called CONNECT. Brokers require from clients, that when sending the CONNECT message, they should define username/password combination before validating the connection, or refusing it in case the authentication was unsuccessful. Overall, security is an ongoing effort for MQTT \cite{Lesjak2015}, and probably the most important one since MQTT is one of the most widely adopted and mature communication protocol solutions. Solving the security issue would create an important and big advantage for MQTT, in comparison with other available solutions.

\subsection{Data Distribution Service (DDS)}
DDS is a real-time data-centric interoperability standard that uses a publish-subscribe interaction model, as defined by the Object Management Group (OMG) \cite{OMG2015}. Unlike some other publish-subscribe protocols, DDS is decentralized and based on peer-to-peer communication, and as such does not depend on the broker component. In DDS, publishers and subscribers can communicate as peers through the data bus, enabling asynchronous data exchange based on their interests. The fact that there is no broker also decreases the probability of system failure because there is no single point of failure for the entire system, making a system more reliable. Both communication sides are decoupled from each other, and a publisher can publish data even if there are no interested subscribers. The data usage is fundamentally anonymous, since the publishers do not enquire about who consumes their data.

One of the salient features of DDS protocol is its scalability, which comes from its support for dynamic discovery. The discovery process, achieved through DDS built-in discovery protocol, allows for subscribers to find out which publishers are present, and to specify the information they are interested in with the defined desired quality of service, and for publishers to publish their data \cite{Corsaro2012}. DDS ensures that proper publish and subscribe nodes will be connected and that the data exchange will be in real-time. Another important and unique characteristic in DDS is its data-centricity, unlike most protocols that are message-centric. For data-centric paradigm what matters the most is the data that clients want to access to, so the focus is on the content information itself. Thus, DDS enables an architecture where participating nodes understand the data value in a consistent manner. In DDS, the data type and content define the communication, whereas in message-centric protocols the focus lies on the operations and mechanisms for delivering that data. The data-centric approach of DDS can be used when system architects define the so-called \emph{topics} by grouping together data items that can logically be related with a goal of ensuring better scalability and performance results.

The main entities in DDS architecture include: Domain, Domain Participant, Topic, publisher, subscriber, Data Writer and a Data Reader \cite{Pardo-castellote}. Publishers and Subscribers are divided into Domains, a virtual concept entity that allows the isolation of communication within nodes that have common interests. Domain Participant is the entry point for message exchange in specific domains, which associates publishers and subscribers and the domains they belong to. It is used to create publishers, subscribers, Data Writers, Data Readers and topics within a domain. The DDS implementation middleware, with data as the main point that will define how the interactions will be conducted, defines how data is structured, changed and accessed in an abstract data space, with a goal of creating a globally shared data \cite{Baptista2001}. The way this is achieved is through a data space abstraction where all the clients can access to read or store their data, known as Global Data Space (\gls{gds}). It is in GDS where the DDS dynamic discovery feature comes into play by allowing publishers and subscribers that join the GDS to automatically discover their mutual existence as well as their interests. The exchange information unit among DDS nodes in GDS is a Topic, and is defined by a name, a data type and a set of QoS policies. Publishers and subscribers are the entities for data distribution and consumption, which publish and receive data through the GDS, but they can not do it on their own. Instead, publishers use Data Writers to send data and subscribers use Data Readers to receive data \cite{Yang2012} with the matching between the two through topics, that is in order to communicate with each other, publishers and subscribers must use the same topic (same name, type and a compatible QoS).

DDS uses UDP by default, but it can also support TCP. Another important protocol in DDS is the Real Time Publish Subscribe (RTPS) \cite{Specification2002} wire protocol, which represents DDS interoperability protocol that allows data sharing among different vendor implementations. One of the advantages of using DDS is a wide set of QoS policies offered (over 20 QoS as defined by the standard). When sending data, the QoS policies of each topic, Data Writers and publishers control how and when the data is sent to the middleware. On the other side, topic QoS, Data Readers and subscribers control the behavior when receiving data.These various policies manage a myriad of DDS features, such as discovery of distributed remote entities, data delivery, data availability, time, and resource utilization \cite{Ingles-Romero2017}.

For a security mechanism DDS implements various solutions. Based on a transport protocol of choice, TLS can be used in case TCP is the transport protocol, or DTLS protocol in case UDP is used.  Similarly for TLS also DTLS brings too much overhead in constrained environments, for which improved mechanisms have been proposed. To this end, the OMG (Object Management Group) DDS Security Specification defines an extensive Security Model and Service Plugin Interface (SPI) architecture designed for for DDS implementations suitable in IoT systems \cite{rtps}. The question of security specification is currently still an open one for DDS and it is expected that new additions will be implemented in the future. One of the additions expected is a secure discovery mechanism  capable of establishing Secure Transport flows between DDS-based applications that have matching security classification, as proposed in \cite{Pradhan2014}.

DDS is an important solution for IoT-based environments for its decentralized publish/subscribe architecture and its support for implementation in both powerful devices and constrained devices \cite{Al-Fuqaha2015}. A challenge of DDS is that it has not been widely used, though this may change with emerging open source DDS implementations ready for testing, such as OpenDDS \cite{OpenDDS}.

\subsection{Advanced Message Queueing Protocol (AMQP)}
AMQP is an open standard protocol that follows the publish-subscribe paradigm as defined by OASIS \cite{OASIS2012}, designed to enable interoperability between a wide range of different applications and systems, regardless of their internal designs. Originally it was developed for business messaging with the idea of offering a non-proprietary solution that can manage a large amount of message exchanges that could happen in a short period of time in a system. This AMQP interoperability feature is significant as it allows different platforms, implemented in different languages, to exchange messages, which maybe especially useful in heterogenous systems \cite{Fernandes2013}.

AMQP has been implemented in two very different versions, AMQP 0.9.1 and AMQP 1.0, each with a completely different messaging paradigm. AMQP 0.9.1 implements the publish-subscribe paradigm, which revolves around two main AMQP entities, both part of an AMQP broker: the exchanges and the message queues. The exchanges represent a part of the broker that is used to direct the messages received from publishers. The publishing of messages to an exchange entity is the first step in the process, and after that messages are routed into one or more appropriate queues. This depends on whether there are more subscribers interested in a particular message, in which case the broker can duplicate the messages and send their copies to multiple queues. A message will stay in the queue until it is received by a subscriber. This routing process, that actually links exchanges and queues, depends on the so-called bindings, which are predefined rules and conditions for message distribution. The newer version of AMQP protocol, AMQP 1.0, is on the other hand not tied to any particular messaging mechanism. While the older versions of the protocol used specifically the above mentioned  publish-subscribe approach with architecture that consists of exchanges and the message queues, new AMQP implementations follow a peer-to-peer paradigm, and can be used without a broker in the middle. Broker is present only in the communication that needs to provide store-and-forward mechanism, while in other cases direct messaging is possible. This option of supporting different topologies increases the flexibility for the possible AMQP based solutions, enabling different communication patterns, such as client-to-client, client-to-broker, and broker-to-broker \cite{IoTprotocols}. It should be noted that a significant amount of infrastructures still use the older AMQP version 0.9.

AMQP uses TCP for reliable transport, and in addition it provides three different levels of QoS, same as MQTT. Finally, the AMQP protocol provides complementary security mechanisms, for data protection by using TLS protocol for encryption, and for authentication by using \gls{sasl} (Simple Authentication and Security Layer).

With all the features it offers, AMQP has relatively high power-, processing- and memory related requirements, making it a rather heavy protocol, which has been its biggest disadvantage in IoT-based ecosystems. This protocol is better suited in the parts of the system that is not bandwidth and latency restricted, with more processing power. 

\subsection{Extensible Messaging and Presence Protocol (XMPP)}

XMPP is an open standard messaging protocol formalized by IETF \cite{Saint-andre2017}, and was initially designed for instant messaging and the exchange of messages between applications. It is a text-based protocol, based on Extensible Markup Language (XML) that implements both client-server and publish-subscribe interaction \cite{Hornsby2010}, running over TCP. In IoT solutions it is designed to allow users to send messages in real time, in addition to managing the presence of the user. XMPP allows instant messaging applications to achieve all basic features, including authentication, end-to-end encryption and compatibility with other protocols \cite{Ramirez2013}.

XMPP supports client-server interaction model, but there are new extensions that enable also for  generic publish-subscribe model to be used. These extensions enable XMPP entities to create topics and publish information; an event notification is then broadcasted to all entities that have subscribed to a specific node. This functionality is rather important for IoT-fog-cloud scenarios, being the foundation for a wide variety of applications that require event notifications. The clients and servers in XMPP communicate with each other using XML streams to exchange data in the form of XML stanzas (semantic structured data units) \cite{Al-Fuqaha2015}. Three types of stanzas are defined: $<$presence/$>$, $<$message/$>$ and $<$iq/$>$ (info/query). A message stanza defines a message title and contents and it is used to send data between XMPP entities. Message stanzas do not receive an acknowledged by the receiving entity, whether it is client or server. A presence stanza shows and notifies entities of status updates, having the role of subscription in XMPP. If there is an interest in the presence of some JID (Jaber ID - a node address in XMPP), a client subscribes to their presence and every time that a node sends a presence update a client will be notified. An iq stanza pairs message senders and receivers. It is used to get some information from the server, for example information about the server or its registered clients, or to apply some settings to the server. Its function is similar to HTTP GET and POST methods.

One of the most important characteristics of this protocol are its security features, which makes it one of the more secure messaging protocols surveyed. Unlike the other protocols surveyed, for example MQTT and CoAP, where the TLS and DTLS encryptions are not built-in within the protocol specifications, XMPP specification already incorporates TLS mechanisms, which provides a reliable mechanism to ensure the confidentiality and data integrity. New additions to the XMPP specifications also include extensions related to security, authentication, privacy and access control. Beside TLS, XMPP implements SASL, which guarantees server validation through an XMPP-specific profile \cite{Conzon2012}.

Since XMPP was initially designed for instant messaging there are some notable potential weakness. By using XML, the size of the messages makes it inconvenient in the networks with bandwidth constraints. Another downside is the absence of reliable QoS guarantees. Because XMPP runs on top of a persistent TCP connection and lacks an efficient binary encoding, it has not been practical for use over lossy, low-power wireless networks often associated with IoT technologies. However, lately, there has been a lot of effort to make XMPP better suited for  IoT \cite{Schuster2014,Che2013,Hornsby2009}. In \cite{Wang2017}, a lightweight XMPP publish/subscribe scheme was presented for resource constrained IoT devices, thus improving and optimizing the existing version of the same protocol.
\\

Throughout this section we saw that one of the main features of all of the above mentioned protocols relevant for their potential utilization and correct placement in an integrated IoT, fog and cloud system is their applicability in resource constrained devices. Table \ref{comparison3} offers a summarized overview of the ongoing efforts of making these protocols more compatible for constrained environments.

\begin{table*}[]%
\caption{Ongoing efforts for constrained environments adaptation}
\label{comparison3}
\begin{center}
\setlength\tabcolsep{3pt}
\begin{tabular}{|c|c|}
  \hline
\textbf{Protocol} & \textbf{Open challenges and efforts in constrained environments}  \\ \hline
REST HTTP     &   TLS version 1.3;  HTTP/2.0 version  \\ \hline
MQTT              &    TLS version 1.3;  MQTT-SN (based on UDP)    \\ \hline
CoAP              & DTLS optimization  \\ \hline
AMQP              &   not recommended for constrained devices\\ \hline
DDS               &   DDS security specification  \\ \hline
XMPP              &     light-weight XMPP publish-subsribe scheme\\ \hline
\end{tabular}
\end{center}
\end{table*}

\section{Performance comparison} \label{section4}

The performance analysis and comparison of communication protocols is still a lively area in the research community. In this section, we survey and analyze the studies reported in different testbed scenarios. An overview of studies focused on performance is shown in table \ref{comparison} and are described in more detail in the following subsections.

\subsection{Latency}
When comparing different parameters for communication protocols, especially for IoT related application, latency comes as one of the priorities. In \cite{Fundam}, authors have analyzed the behavior of two HTTP and MQTT in a fog-to-cloud IoT based architecture scenarios. The results of the experiments have shown that the measured response times for the requests were shorter for MQTT then the ones for HTTP.
In \cite{Joshi2017}, a Raspberry Pi based home automation system was used along with a web server and smart phones to measure latency generated by the MQTT (Mosquitto) and HTTP (REST) based architectures. As a result,  MQTT-based architecture produced lower latency. In \cite{Thangavel2014}, it was shown that MQTT messages had experienced lower delays than CoAP for lower packet loss and higher delays than CoAP for higher packet loss. The comparison of these two protocols has also been conducted in \cite{DeCaro2013} where authors assessed latency by measuring RTT. The results have shown that the average CoAP RTT was more than 20\% shorter than MQTT. Another comparison of RTT in MQTT and CoAP \cite{Mijovic2016} was conducted in two scenarios, local area network and an IoT network, with average RTT being from two to three times higher in the IoT network scenario. The results showed that MQTT with QoS0 had lower RTT in comparison with CoAP, while MQTT with QoS1 had the higher RTT due to the presence of both transport and application layer ACKs. In \cite{Iglesias-Urkia2017}, the latency of MQTT and CoAP was analyzed for different QoS levels in a network without congestion. These conditions favored CoAP because it required fewer bytes to transfer the same message with a shorter delay regardless of the QoS level. Regarding the latency, MQTT latencies were measured in the order of milliseconds,  and CoAP latencies as low as hundreds of microseconds. However, it is important to notice that in the cases of less reliable networks, MQTT\'s underlying TCP protocol will be an important advantage and the results would be different.

The latency comparison in \cite{IoTprotocols} for the two broker based protocols, AMQP and MQTT, for increasing payload size showed that when transferring relatively small payloads the latencies of the two protocols are almost the same, but when transferring huge payloads MQTT yields a lower latency.

In \cite{Dimcic}, CoAP was compared with HTTP in a machine-to-machine communication scenario with devices deployed on a top of the vehicles and equipped with the gas sensors, weather sensors), location (GPS) and a mobile network interface (GPRS). The time needed to transfer a CoAP message over mobile network was almost three times shorter then the time required when HTTP messages are used. Another comparison of these two protocols was conducted in \cite{Saleh2016} as potential communication protocols for smart grid devices over the Arduino hardware platform. The results have shown the HTTP had a longer response time. An emulation-based quantitative performance assessment of CoAP in comparison with HTTP was conducted in \cite{Gao2017}, taking into account different QoS levels of CoAP (with confirmable and non-confirmable messages). Again, HTTP showed a comparably poor delay performance.

Including more protocols in their analysis, the authors in \cite{Chen2016} compared the performance of IoT protocols MQTT, DDS and CoAP in a medical application scenario using a network emulator. DDS outperformed MQTT in terms of experienced telemetry latency in various poor network conditions. The UDP based CoAP performed well for the applications that required low latency; however, since it is UDP based there was a significant amount of unpredictable packet loss. In \cite{Babovic2016}, authors compared web performance of publish/subscribe IoT messaging protocols MQTT, AMQP, XMPP, and DDS by measuring the latency of sensor data message delivery and the message throughput rate. The results have to be taken with a reservation because they heavily depended on the message broker and JavaScript client implementations. The shortest latency was produced by the MQTT protocol, followed by AMQP, while the difference between XMPP and DDS was negligible.  Authors in \cite{Naik2017} compared MQTT, CoAP, HTTP and AMQP messaging protocols based on their average latency among other parameters. The results have shown the highest latency in HTTP, followed by AMQP and MQTT respectively, with CoAP having the  lowest latency results. Finally, it should be noted that only a few papers compare a new HTTP version, HTTP2/0 with the other messaging protocols, or evaluate HTTP2/0 performances in IoT scenarios. The paper \cite{Daniel2014} compares the IoT adapted \gls{spdy} (Speedy) protocol, which was used a basis for HTTP/2 with CoAP and HTTP. The experiments showed that CoAP has the lowest download time and the least number of bytes transferred. In \cite{Londono2016}, the data transfer time for CoAP and HTTP/2 was compared, with HTTP/2 having better results in high congestion scenarios and CoAP in lower congestion scenarios.

To summarize, even though there is no comparative study of all the protocols discussed here, we can conclude that the latency is heavily influenced by  the underlying transport protocol, and the use of TCP in MQTT, AMQP, HTTP and XMPP is a major factor that causes higher latency values than in CoAP and UDP based DDS.

\subsection{Bandwidth consumption and throughput}
In \cite{Thangavel2014} MQTT and CoAP have been analysed using the common middleware in terms of bandwidth consumption that was measured as total data transferred per message. In the cases where message size was small, and independently of the increase of packet loss rate, CoAP consumed less bandwidth than MQTT. The authors in \cite{Fernandes2013} calculated protocol efficiency, as as the ratio between the number of useful information bytes and the total number of bytes exchanged at application and transport layers, and used it to compare MQTT and CoAP. The results showed higher efficiency for CoAP.
An emulation-based quantitative performance assessment of CoAP in comparison with HTTP was conducted in \cite{Gao2017} in the dynamic network environment. This scenario included a large amount of devices transferring data at the same time, which is a typical case in IoT environments. In order to achieve a higher utilization results have shown that it is better to use CoAP.
Alongside latency, the authors in \cite{Chen2016} compared bandwidth consumption for three protocols MQTT, DDS (with TCP as a transport protocol) and CoAP once as a function of a network packet loss and once as A function of network latency. CoAP generally showed a comparably lower bandwidth consumption that did not increase with increased network packet loss or increased network latency, unlike MQTT and DDS, where bandwidth consumption increased in mentioned scenarios. Also DDS consumed approximately twice the bandwidth of MQTT. In another study, \cite{Tandale}, authors have also compared three protocols, this time MQTT (taking into consideration all three QoS level options), CoAP and REST HTTP in terms of bandwidth measurements. The scenario in question covered IoT to cloud communication and the results heavily depended on the size of the payloads that were being transferred. In the case of small payloads CoAP used the least amount of bandwidth, followed by MQTT and REST HTTP. However, when the size of payloads increased, the best performances were measured for REST HTTP.

\subsection{Energy consumption}
The power/energy consumption is essential in every IoT based system, and the choice of protocols affects the same. In \cite{Joshi2017} along with latency analysis, authors have compared the energy consumption between MQTT and HTTP, with the results that energy consumed by HTTP was much larger than with MQTT. In \cite{Bandyopadhyay2013} authors have analysed average energy consumed by MQTT and CoAP for a constrained gateway device with experimental results showing that CoAP is more efficient in terms of energy, though both of them proved to be efficient. A similar conclusion can be found in \cite{Thota2017} where authors have shown that in the simple scenarios MQTT was more suitable for IoT messaging and nodes with no power constraints. CoAP on the other hand has proved to have efficient power management capabilities. In \cite{Colitti2011} authors provides an evaluation of CoAP compared to HTTP which demonstrated that thanks to the smaller headed and packet size in CoAP results in it having a lower energy consumption. In \cite{Luzuriaga2015b} authors compared the capabilities of AMQP and MQTT under a mobile or unstable wireless network testbed with the conclusion that AMQP offered more aspects related to security and MQTT was more energy efficient. Similar to other performance measures presented in this survey, most of the papers compare a pair of protocols in one study, and no study has evaluated and compared the energy consumption of all candidate communication protocols.

\subsection{Security}

Security remains one of the most important challenges as pointed out by a large number of  research papers \cite{Kumar2015, Granjal2015, FRUSTACI2017, Jing2014, Tiburski2015, Fremantle2015, Dragomir2016, Nastase2017}.  Related to security, we focus here on application layer, where it is also necessary to understand the communication challenges related to performance factors such as latency, overhead and packet loss. As mentioned in previous sections, the choice for security mechanism for surveyed protocols is usually based on TLS or DTLS protocol with protocols like HTTP, MQTT, AMQP and XMPP basing their security on TLS, CoAP on DTLS, and DDS supports both options. Both TLS and DTLS start with the handshaking process between client and server side in order to exchange supported cipher suits and keys, and agree on the ones both sides support to assure that further communication happens in a secure communication channel. The difference between the two is in small modifications that allow UDP-based DTLS to run over an unreliable connection. The slight advantage of TLS is that it is a widely used and stable security protocol, with a software client and server support and in available cryptography libraries \cite{Kumar2015}. In \cite{AlFardan2013} authors presented distinguishing and plaintext recovery attacks against TLS and DTLS with experimental results demonstrating the feasibility of the attacks in realistic network environments for several different implementations of TLS and DTLS. The results reported were in favor of TLS since the attacks proved to be much more serious for DTLS, because of its tolerance of errors.

However, the biggest issue with the implementation of these protocols in IoT-F2C systems is that they were not originally designed for utilization in IoT and constrained devices. Through their handshake process, they add additional traffic with each connection establishment that drains the computing resources. In \cite{Tiburski2017} the use of TLS and DTLS protocols in communication channel was analysed and compared with their corresponding insecure options, TLS communication was compared with regular TCP-based exchange and DTLS communication with regular UDP-based exchange. On average results showed an increase of 6.5\% for TLS and 11\% for DTLS in overhead, compared to communication without security layer. In resource rich environments that are usually located in the cloud layer, this would not be a problem, but in the IoT to fog layer communication this becomes an important limitation. For these reasons, security is an ongoing effort with a goal of optimizing TLS and DTLS by creating more lightweight versions, or finding alternative solutions.

\subsection{Developer's choice}
One important factor for any protocol adoption is the choice by system developers. The adoption of the protocols presented here was in fact surveyed collaboratively by  Eclipse IoT Working Group, IEEE, Agile-IoT EU and the IoT Council in order to better understand how developers are building IoT solutions \cite{EclipseIoTWorkingGroup2016}. On the question what messaging protocol is used in IoT solutions, the results of that analysis have shown that MQTT and HTTP are the most used and adopted protocols. The reason for this is that MQTT and HTTP REST are currently  comparably more mature and more stable IoT standards than other protocols. For many IoT developers, MQTT and HTTP are protocols of choice in their IoT, fog and cloud implementations.

\section{Connecting the IoT, Fog and the Cloud} \label{overview}
We now focus on the communication protocols and discuss their positioning within a combined IoT-fog-cloud architecture, which per se is an open direction for future research. It should be first noted that no communication protocol has been originally designed for a combined IoT-fog-cloud systems, and there is no unifying standard, and this alone is an ongoing area of research and development. On the other hand, all of the above mentioned protocols are operating in the application layer of the OSI protocol stack, and as such they could, in theory, fill the same role in various parts of the system. This section is dedicated to discussing the pros and cons of single- or multiple protocol solutions in various segments, opening this area for broader dissemination and further reserach. We conclude this section with a summarized discussion of other open issues and challenges.

\begin{figure}[t]
	\centering
	\includegraphics[width=0.6\linewidth]{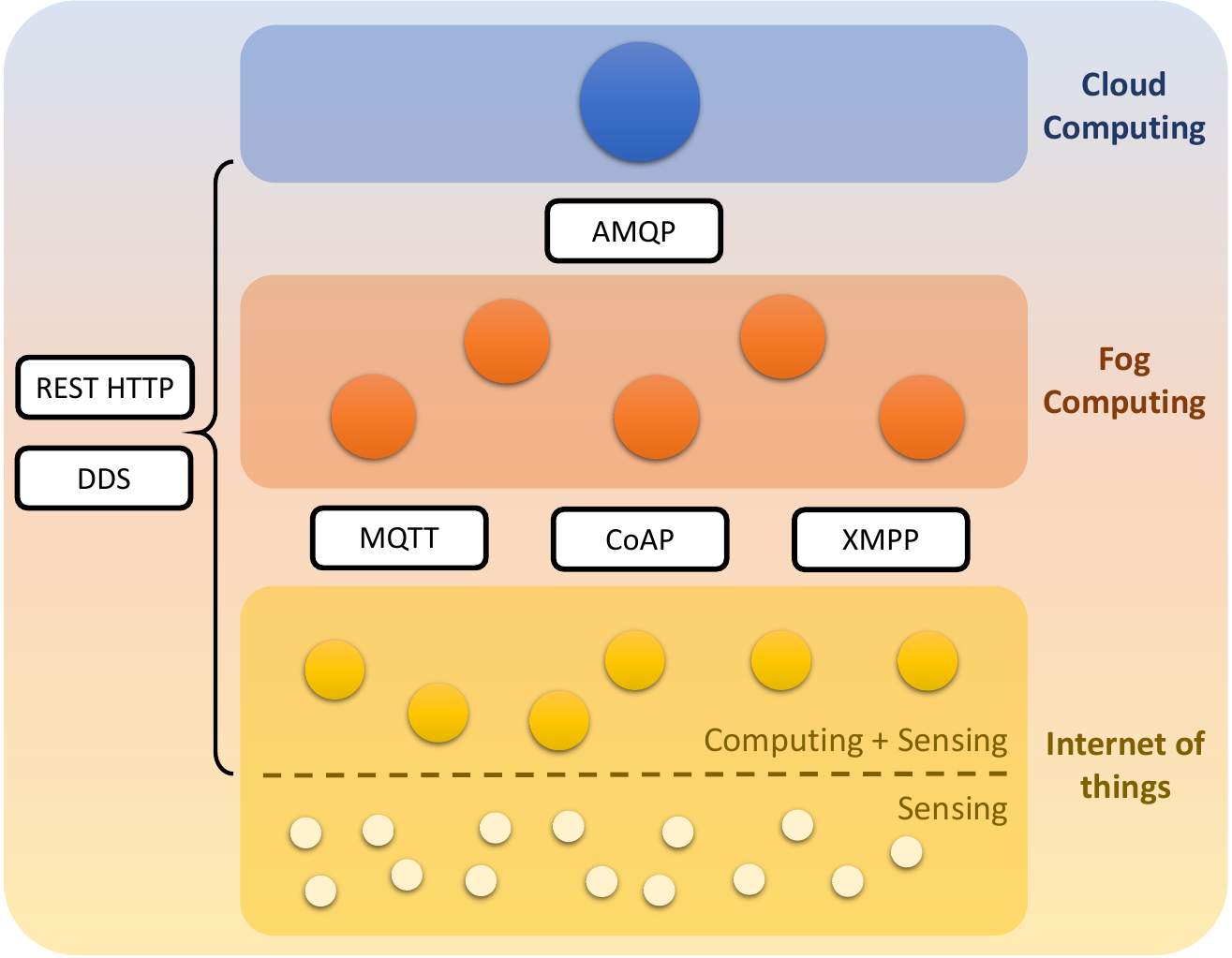}
    \caption{Application protocols in IoT, Fog and Cloud networks}
    \label{arch2}
\end{figure}

\subsection{Protocol solutions based on a single communication protocol}
Let us start the discussion with an example. We first note that some of the protocols previously described are more widely used and accepted than the others, such as MQTT and REST HTTP, and are candidates for a  single-protocol solutions. The practice of single-protocols solution has many downsides, however, since based on the features considered,  it is obvious that each of the protocols can optimally satisfy specific requirements. For example, a protocol that satisfies constrained environment can underperform in the domain that has strict security requirements. With this in mind, we consider two candidate single protocol based solutions in combined IoT, fog and cloud, and these are MQTT and REST HTTP. It should be mentioned that based on its characteristics for this kind of a solution DDS could also be considered, and in fact Vortex DDS Platform offers DDS based solutions both in cloud and fog based IoT systems. However, at this moment, MQTT and REST HTTP based solutions are much more widely accepted and as such are of the more interest here.

\subsubsection{REST HTTP as a single protocol solution}
\begin{figure}[t]
	\centering
	\includegraphics[width=0.7\linewidth]{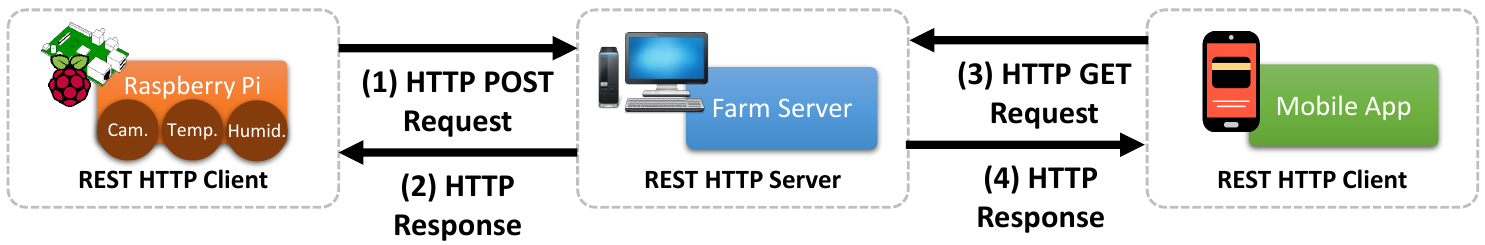}
    \caption{REST HTTP request/reply interaction model}
    \label{rest_example}
\end{figure}

\begin{figure}[t]
	\centering
	\includegraphics[width=1.0\linewidth]{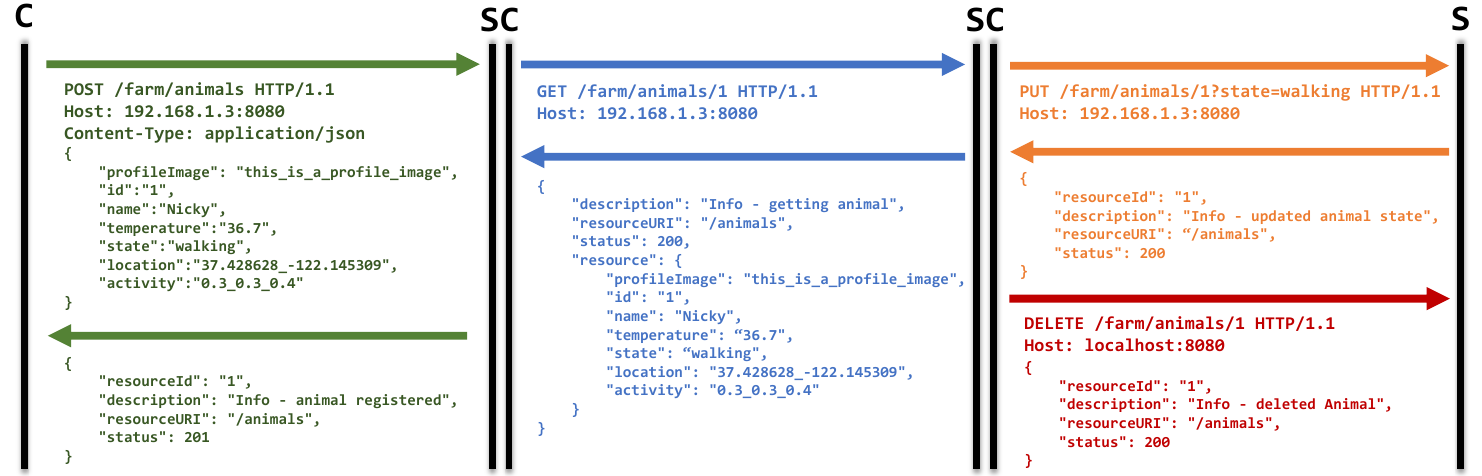}
    \caption{Example of REST HTTP methods for CRUD operations; C: Client, S: Server)}
    \label{rest_http_crud}
\end{figure}

\begin{figure}[t]
	\centering
	\includegraphics[width=0.7\linewidth]{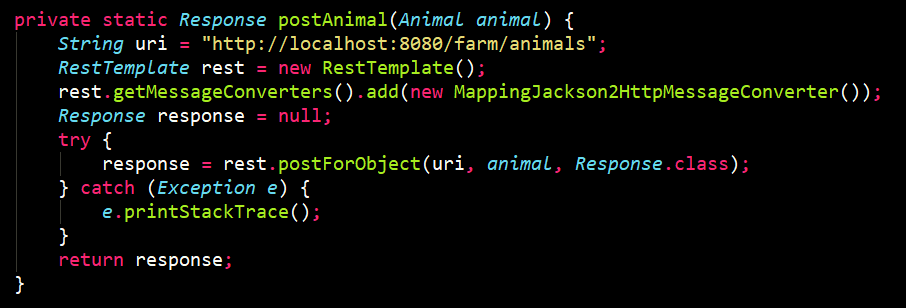}
    \caption{Example of a HTTP Post request in Java using Spring Framework}
    \label{REST_HTTP_code}
\end{figure}

Fig. \ref{rest_example} illustrates REST HTTP request/reply interaction in an IoT-to-fog application in smart farming, which we adopted from \cite{Carpio2017}. A code example of how HTTP methods support CRUD operations in this kind of applications shown in Fig. \ref{rest_http_crud}. In this example, animals are equipped with wearable sensors (IoT Client, C) and managed in a fog computing smart farming system (fog server, S). Here, in the header of the POST method the resource to modify is specified \emph{/farm/animals}, as well as HTTP version and \emph{Content-type} which in this case is a JSON object that represents a farm animal to be managed by the system (Nicky, the cow). In this example the response from the fog server specifies that the request was successful, with the HTTPS status code  \emph{201, resource created}. The GET method only needs to specify the requested resource in the URI (e.g. \emph{/farm/animals/1}), which returns, in this example, the JSON representation of the animal with this id from the server. Similarly, the PUT method is used when some specific resource entry needs to be updated, in this case, in the resource URI is specified for the parameter to be changed and the current value (e.g., for instance indicating that cow is currently walking, \emph{/farm/animals/1?state=walking}). Finally, the DELETE method is used equally than the GET method, but just deleting the resource as a result of the operation. As mentioned in Section \ref{http_subs}, REST HTTP has many available frameworks which makes its utilization an easy and logical choice. For instance, for Java developers, Spring Framework \cite{SpringRest} facilitates the implementation of RESTful web services. We illustrate an example in Fig. \ref{REST_HTTP_code}. This method creates an HTTP POST request to the specified URI parsing an (animal) Java object into a JSON format. The response object maps the response from the server to a Java object in order to be managed by the application. \\

 \begin{figure}[t]
 	\centering
	\includegraphics[width=0.7\linewidth]{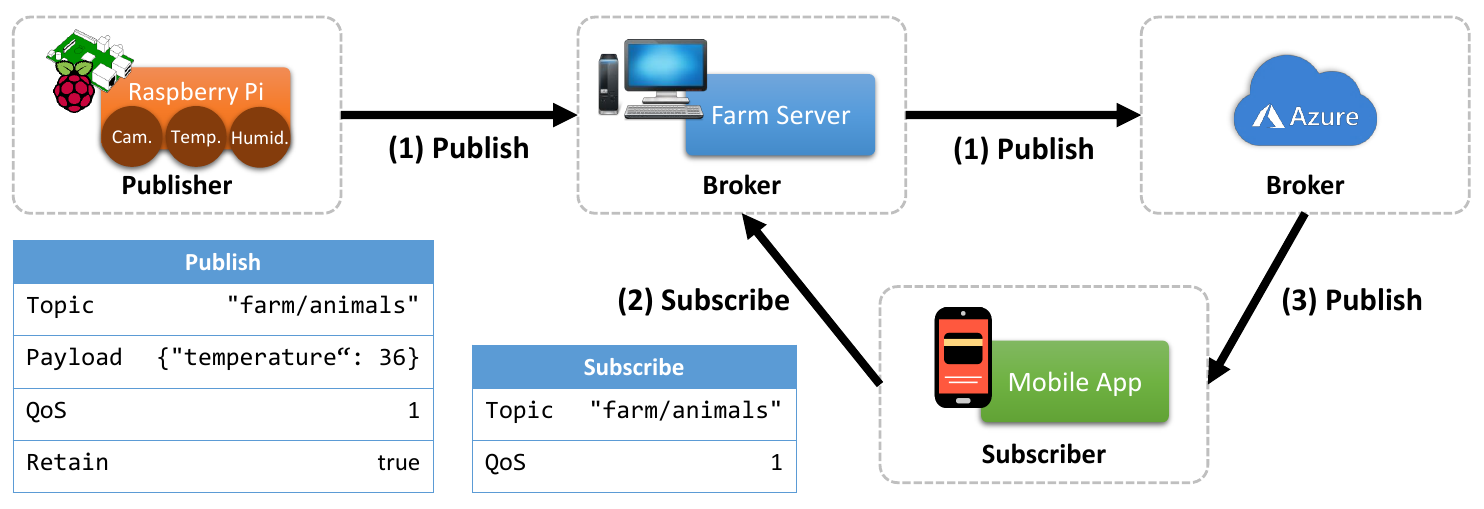}
 \caption{Connecting the IoT, Fog and the Cloud - MQTT example}
  \label{mqtt_example}
 \end{figure}
\subsubsection{MQTT as a single protocol solution}
Let us use the same example of a smart farm, but in this case for the communication instead of REST HTTP, MQTT protocol is used, as illustrated in Fig. \ref{mqtt_example}. Let us first describe IoT to fog communication, and then extend it to the third abstract layer - the cloud. The local server with the installed Mosquitto library has a role of the broker, in this example a simple off the shelf personal computer (denoted as farm server). A Raspberry Pi serves as a MQTT client, realized by installing MQTT Paho Library that is fully compatible with the Mosquitto broker. This client corresponds to the IoT abstraction layer, representing a device with sensing and computing capabilities. The broker, on the other hand, corresponds to the higher abstraction layer representing a fog computing node, characterized by larger computing and storage capacities. In the proposed smart farm scenario Raspberry Pi is connected to accelerometer, GPS and temperature sensors and publishes data from these sensors to a broker fog node. As explained Section  \ref{mqtt_subs} topics in MQTT are treated as a hierarchy. One MQTT publisher can publish messages to a defined set of topics, in this case, three topics. For the sensor that measures the temperature in an animal shed, a client will publish the temperature under the following topic: {\em animalfarm/shed/temperature}. In the same manner, for the sensors that measure GPS location and animal movement through accelerometer, a client will publish the corresponding updates under the following topics: {\em animalfarm/animal/GPS} and {\em animalfarm/animal/movement}. This information will be published to the broker which can temporally store it in a local database in case that later another interested subscriber appears.

\par In addition to a local server that has a role of a fog MQTT broker to which Raspberry Pis that serve as MQTT clients publish data from the sensors, there can be another MQTT broker in the cloud layer. In this case, the information published to the local broker can temporally be stored in a local database and/or transmitted to the cloud. Here, the fog MQTT broker is actually used to bridge all the data to another MQTT broker that is represented by a cloud based instance. With such an architecture, the user with a mobile application can be subscribed to both brokers. In this way, if connection with one of the broker fails, say cloud, the end user has the option of receiving the information from the other one, e.g. fog. This is a salient feature of combined fog and cloud computing systems. By default, mobile application can be configured to first connect to the fog MQTT broker, and if not successful, to connect to the cloud MQTT broker. This MQTT bridging solution between one local broker serving as a fog node and one cloud broker is just one of the possible solutions in IoT-F2C systems. Related work so far has more commonly considered MQTT from the IoT device layer to fog/edge nodes, while the communication from fog to cloud was left to other candidate application layer protocols \cite{Peralta}. This leads us to the solution with multiple communication protocols as we will discuss next.

\begin{figure}[t]
	\centering
	\includegraphics[width=0.7\linewidth]{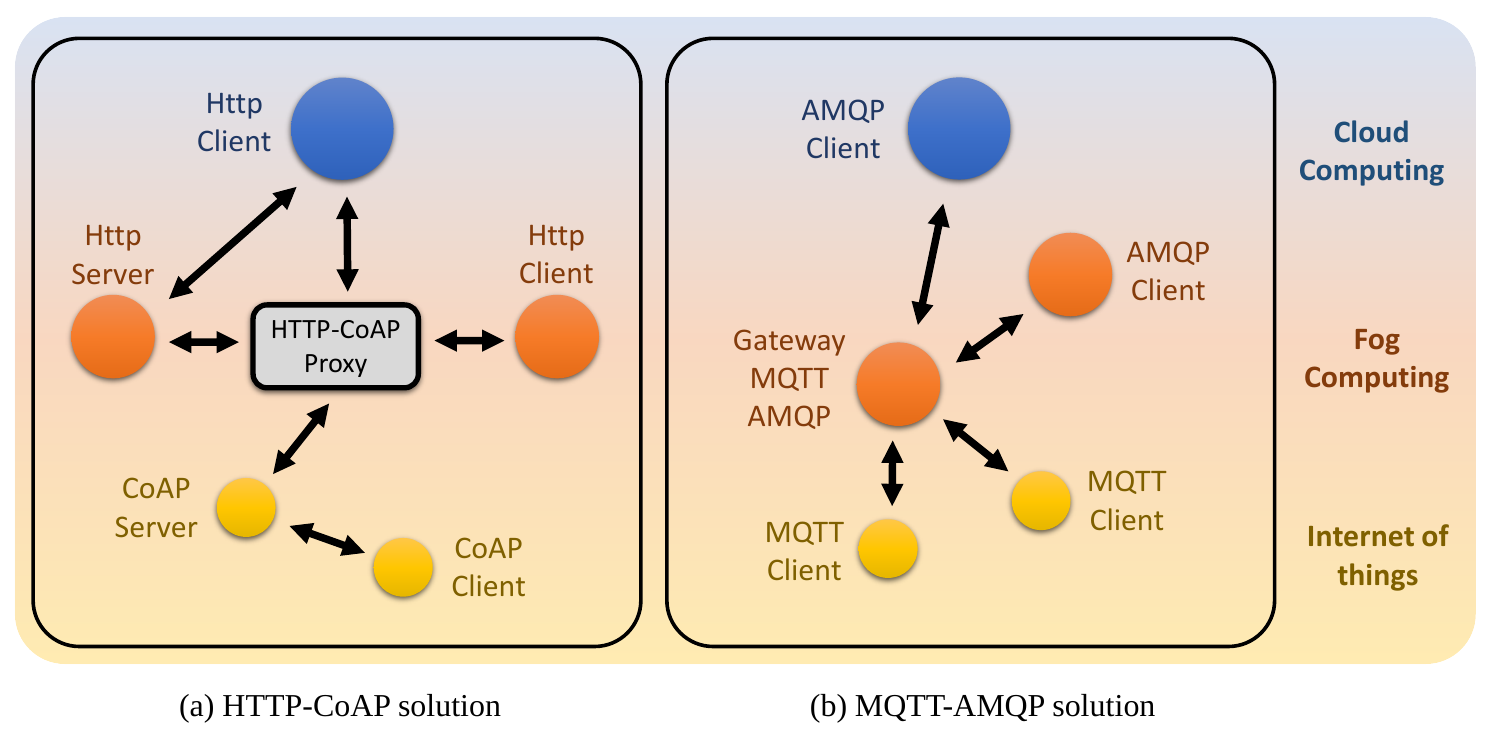}
   \caption{Connecting the IoT, Fog and the Cloud}
    \label{iot_fog_cloud_solutions}
\end{figure}

\subsection{Multiple protocol solutions based on a combination of communication protocols} \label{multiplepr}
While the single protocol solutions have been popular because of their easier implementation, it is obvious that in IoT-F2C systems it would make sense to combine different protocols. One of the findings of this survey is in fact that individual protocols can be better positioned within parts of the overall system, as illustrated in Fig. \ref{arch2}. Let us consider, for illustration, three abstraction layers of IoT, fog and cloud computing. The devices in the IoT layer are generally considered as constrained. For the sake of this survey, let us consider the IoT layers as the most constrained, cloud the least constrained and fog computing as "somewhere in between." Based on this assumption, we find and Fig. \ref{arch2} shows that between the IoT and the fog abstractions, the current protocol solutions include MQTT, CoAP and XMPP. Between the fog and the cloud, on the other hand, AMQP is one of the main protocols used, together with REST HTTP, which due to its flexibility is also used between IoT and fog layers. At the first sight, however, RESTful HTTP protocol and the newly proposed DDS protocol can be used in all layers. We will get back to this and other observations in the following sections, with more in-depth discussions on single- and multiple protocol solutions.

For solving the communication among two lower abstract layers, that is IoT and fog, the common denominator that determines the suitability of a protocol is the ability to run as a lightweight protocol on constrained devices. On the other hand, this requirement is not necessary for the communication among fog and cloud layer. Based on their characteristics and the scenarios we encountered during our survey, the straightforward solution would include the combination of lightweight protocol between IoT and the fog and a protocol not restricted to the constrained devices between the fog and the cloud. The main issue with these kind of solutions, however, is protocol interoperability and the ease of translating the communication from one protocol to another, presenting also a challenge for recent research efforts \cite{Desai}. Ideally, in the future, an IoT-F2C system architecture will be independent on the communication protocol used, and will provide integration among different protocols. Since this is not the case at the moment, in order to avoid additional implementation difficulties it makes more sense to combine protocols without significant conceptual differences. To this end, one potential solution is based on the combination of two protocols that follow the same architectural style, REST HTTP and CoAP. The other proposed solution is based on the combination of two protocols that follow the publish-subscribe interaction, MQTT and AMQP. Following the similar concept (both MQTT and AMQP are broker based, CoAP and HTTP both use REST style) makes these combination easier to implement, and require less integration efforts. It should be noted that other combinations are also possible, such as MQTT and REST HTTP, but are more difficult for realization.

\subsubsection{REST HTTP-CoAP example}
Fig. \ref{iot_fog_cloud_solutions} (a) shows the two request-reply based models, HTTP and CoAP, and their possible placement in an IoT-F2C solution. Since HTTP is one of the best known and adapted protocols in current networks, it is unlikely that it will be completely replaced with other messaging protocols. Among the nodes that present powerful devices, which would be between cloud and fog, REST HTTP is a reasonable solution. On the other hand, for resource constrained devices that communicate between fog and IoT layer it is more efficient to use CoAP. One of the big advantages of CoAP is in fact in its interoperability with HTTP, being that the both protocols are based on REST principles. For this interoperability to work it is necessary to deploy a proxy between them that will allow HTTP clients to request resources from CoAP servers and CoAP clients to request resources from HTTP servers \cite{Lerche2012} also presented in Fig. \ref{iot_fog_cloud_solutions} (a). The reference information for implementing a proxy that performs translation between HTTP and CoAP is given in an document by CoRE working group \cite{Fossati2015}. A lot of research effort is put into developing and analysing HTTP-CoAP proxies and mappings between the two \cite{VanDenAbeele2017,Esquiagola2017,Sulaeman2016,LeSommer,Castellani2012, Buschsieweke2017}. \\

\subsubsection{MQTT-AMQP example}
Fig. \ref{iot_fog_cloud_solutions} (b) alternatively shows two publish-subscribe interaction based models in the same scenario, including MQTT and AMQP. While hypothetically, both protocols could be used for communication among nodes in every abstraction layer, their position should be decided based on the performance. MQTT was built as a lightweight protocol for devices with limited resources so it could be used for communication between IoT constrained nodes and fog nodes. AMQP is also a lightweight protocol; however, with additional support for security, reliability, provisioning and interoperability the overhead and message size also increase, thus degrading its performances in nodes with limited processing power. For these reasons, AMQP is more suitable in the more powerful devices, which would position it ideally between fog and cloud nodes. Instead of MQTT in IoT to cloud domain, based on the fact that is considered lightweight and as such adjusted for constrained devices, it is possible to use XMPP protocol. But, at this moment, similar to DDS, it is not as widely accepted in this kind of scenarios.

\begin{table}%
\centering
\caption{Performance comparisons related to application layer protocols}
\label{comparison}
\begin{minipage}{\columnwidth}
\begin{center}

\setlength\tabcolsep{4pt}
\begin{tabular}{|c|c|c|c|c|c|}
\hline
 &  & \textbf{Bandwidth} & \textbf{Energy} &  & \textbf{Developer's}    \\
 \textbf{Protocol} & \textbf{Latency} & \textbf{utilization} & \textbf{consumption} & \textbf{Security} & \textbf{choice}    \\
 &  & \textbf{and throughput} &  &  &    \\ \hline
REST HTTP   &  \cite{Joshi2017, Dimcic, Saleh2016, Fundam}  &  \cite{Gao2017, Tandale}   &  \cite{Joshi2017, Colitti2011}   &  \cite{7958594, 7546519}  &  \checkmark    \\
    &   \cite{Gao2017, Naik2017, Daniel2014, Fundam}   &     &    &   &\\ \hline
  &  \cite{Joshi2017, Thangavel2014, DeCaro2013}   &        &      &    &   \\
 MQTT    & \cite{Mijovic2016, Iglesias-Urkia2017, IoTprotocols}   &   \cite{Thangavel2014, Fernandes2013, Chen2016, Tandale}    &   \cite{Joshi2017, Bandyopadhyay2013, Thota2017, Luzuriaga2015b} & \cite{Lesjak2015}  & \checkmark \\
 &  \cite{Chen2016, Babovic2016, Naik2017}    &     &    &   &  \\ \hline
     &  \cite{Thangavel2014, DeCaro2013, Mijovic2016, Iglesias-Urkia2017}   &    &   & \cite{Kumar2015, Raza2013, 6227754}   &   \\
CoAP  &  \cite{Dimcic, Saleh2016, Gao2017, Chen2016}   &   \cite{Thangavel2014, Fernandes2013, Gao2017, Tandale}   &  \cite{Bandyopadhyay2013, Thota2017, Colitti2011}  &   \cite{Granjal2015, 7103240}    &   \\
  &  \cite{Naik2017, Daniel2014, Londono2016}   &      &   &     &   \\ \hline
AMQP   &  \cite{IoTprotocols, Babovic2016, Naik2017}   &  -   &  \cite{Luzuriaga2015b}   &  - &             \\ \hline
DDS   &  \cite{Chen2016, Babovic2016}    &  \cite{Chen2016}    &  -   &  \cite{rtps, Pradhan2014}    &   \\ \hline
XMPP     & \cite{Babovic2016}      &   -   & -   & \cite{Conzon2012}    &   \\ \hline
HTTP/2.0       & \cite{Daniel2014, Londono2016}     &   -  &  -     & -    &     \\ \hline
\end{tabular}
\end{center}
\end{minipage}
\end{table}

\subsection{Open Issues and Challenges Summarized}

Until now, with the IoT advancements, a significant amount of the research efforts has been put into the comprehensive comparisons of the different communication protocols that could potentially be used in the IoT related applications and scenarios. These efforts include comparisons based on the different characteristics of these protocols (underlying transport protocol, interaction model, security, quality of service) as well as based on their individual performance strengths and weaknesses in different IoT related systems.

There is overall a lack of a comparative study of all of the mentioned protocols in a scenario that would cover a broader architectural paradigm that combines IoT, fog and cloud computing systems, leaving it as a grand challenge for a future research. The next step towards this goal would be to evaluate performance of various protocols surveyed in a useful application scenario.  This should include evaluation of the communication in IoT-F2C when each of the individual protocols is used, as well as the scenarios when the two, or more of these protocols are used at the same time in the system.  Using multiple communication protocols solutions in fact presents an important new research direction, their interoperability and interaction models. As mentioned in the Section \ref{multiplepr} there is still an open issue of combining a publish-subscribe and client-server communication between different parts of IoT-F2C. For some of the protocols, such as DDS, which follows completely different architecture from the others, it us unlikely that is even possible to use it in a combination with other protocols, which is an open issue.

While there are many comprehensive studies on different protocol parameters, the most important one that is constantly being adapted for IoT purposes is the security, and it should be privacy as well. Some of these aspects can be addressed by using the TLS and DTLS in combination with surveyed communication protocols, though by doing so they would lose their lightweight properties. The research on how to adapt these two security mechanisms is still ongoing. The area of privacy remains generally under-addressed, including aspects of anonymous communication and censorship applications. While some papers \cite{FRUSTACI2017, Marian, Huang} tackle general IoT or specific protocol privacy issues the area of privacy still requires major efforts to advance, both in application layer which was the subject of this layer, and in IoT-F2C solutions.


Finally, a straightforward next future direction need to consider recent developments in newly proposed protocols, most notably HTTP 2.0 and QUIC, that with no doubt will leave their mark towards seamless integration and coordination of IoT, fog and cloud computing systems.

\section{Conclusions and Outlook}

We surveyed application layer protocols designed or adapted for utilization in IoT solutions, focusing on their possible implementation in the IoT-based fog and cloud computing systems. For a system that has to take into account different requirements for IoT, fog computing and cloud computing, it is not likely that any of the surveyed protocols alone will be enough to cover the entire communication in the system, starting from resource constrained devices over to the cloud servers. The survey found that the two most mature choices to consider, which also are favored by developers, to be MQTT and RESTful HTTP. These two protocols are not only the most mature and stable ones, but also include many well documented and successful implementations and online resources. Based on its stability and simple configuration MQTT is the protocol that has proven over time to have excellent performance when used in IoT layer with constrained devices. In the parts of the system where the constrained communication and battery consumption are not an issue, such in some fog and most cloud computing systems, RESTful HTTP is a straightforward choice. CoAP should also be taken into consideration as it is also rapidly evolving as an IoT messaging standard and it is likely that in the near future it will reach a level of stability and maturity similar to MQTT and HTTP. But the standard is evolving for now, which carries short-term interoperability challenges.

One of the major challenges we identified, which is among key factors when choosing appropriate protocols, is that of defining standards to unify varying architectures and interfaces with a goal of achieving a combined management of IoT, cloud and fog. While there are architecture and system proposals that offer IoT-cloud based solution, integrating them with fog computing paradigm is still a novel proposition. Because of the different architectural possibilities, each protocol performs differently in different segments and is thus suited for different types of applications. One of the challenges is also related to the usage of  proprietary protocols and their interoperability with increasingly important open protocols. Another issue is the one of implementation stability, as mentioned above, which is a key factor for protocol choice for system developers. Implementing application protocols other than HTTP requires training for developer's teams. Performance studies have shown that the REST HTTP is however not sufficient in combined IoT, fog and cloud solutions, and this is an open issue for reserach. It remains to be seen whether the future protocol choice will include other messaging protocols, or focus on improving HTTP as it is. Other important features, such as security and privacy, need to be also further analyzed on the overhead they bring, as the current solutions are far from optimal. This creates not only challenges but also exciting opportunities in novel architectures that without doubt will need to combine IoT, fog and cloud computing systems to meet the requirements of future applications.

\section*{Acknowledgment}
 This work has been partially performed in the framework of mF2C project funded by the European Union's H2020 research and innovation programme under grant agreement 730929.

 \clearpage
\printnoidxglossary[nonumberlist,style=super,nogroupskip]

 \clearpage
\bibliographystyle{unsrt}

\bibliography{survey-bibliography}

\end{document}